\documentclass[11pt]{article}

\newcommand{\beq}{\begin{equation}} 
\newcommand{\eeq}{\end{equation}}

\begin{document}

\title{\large \bf The Missing Mass of the Milky Way Galaxy}
\author{Kenneth Dalton  \\
                        \\
        e-mail:  kxdalton@yahoo.com}
\date{}
\maketitle

\begin{abstract}
This work verifies the principle  that  
gravitation is caused by {\it energy} and not simply by mass alone.  
A model is proposed in which cosmic ray protons flow
radially through the galaxy.  The resulting electric field energy creates a
gravitational force, in addition to the conventional Newtonian force.  The model yields a rotation
curve that agrees in detail with the experimental curve.  The total electric
field energy is calculated.  It is the missing mass of the galaxy.

%Keywords: Galaxy: kinematics and dynamics; rotation curve; gravitation
\end{abstract}

\newpage

\section*{\large {\bf 1. Introduction.}}

In the proposed model, electrons are separated from protons by the galactic
magnetic field.  Similar charge separations are known to occur in our solar
system.  For example, the solar wind is a plasma with 
enough thermal energy to escape the Sun.  The flux is neutral, composed 
almost equally of protons and electrons (there are some ions, as well).
The total flux is $10^{36}$ particles $\mbox{\rm sec}^{-1}$, moving with
an average velocity of $400$ km $ \mbox{\rm sec}^{-1}$.  However,
the lighter electrons have the greater outward velocity, and a charge 
separation occurs.  This gives rise to an electric field which aids the 
acceleration of ions. [1]

A second example is provided by cosmic rays, which are very energetic
charged particles.  The cosmic ray flux arriving at Earth is not neutral: protons make
up 90\% of the total.  It is thought that magnetic fields in the solar
system deflect the electrons, thereby causing the observed separation
of charge.  

The central bulge of the Milky Way Galaxy contains some $10^{10}$ stars.
Assuming a thermal flux similar to that of the Sun, one would expect
the bulge to emit $10^{46}$ particles $\mbox{\rm sec}^{-1}$.   
The magnetic field in the disk preferentially separates electrons.
They flow outwardly through the disk, following the magnetic lines.  
The free protons are mutually repelled and move rapidly away from each other,  
creating a radial electric field in the galaxy.
The electric field is such that the protons reach cosmic
ray energy.  It is found that the density of free protons is very
small, about 1 $\mbox{\rm km}^{-3}$.  Nevertheless, the energy of
the electric field produces a gravitational force, which is
greater than that produced by the mass of stars alone.

%\newpage
\section*{\large {\bf 2. The Electric Field.}}

The rotation curves of many galaxies have now been measured. They exhibit a 
characteristic flat velocity profile at points far from the galactic center.
These observations cannot be understood in terms of conventional Newtonian theory, 
if visible mass is taken to be the sole source of gravitation.  This apparent failure 
underlies the belief that a great deal of mass is missing from the cosmos.  

Here, the argument is made that electric field energy contributes to the gravitational
field of the galaxy.  The electric field arises from cosmic ray protons that emanate from 
the central bulge.  They stream radially through the core, through the halo,
and eventually exit the galaxy.   
These protons are confined only by the limiting speed of light.  
In a steady state, the amount of proton charge passing
through each spherical surface is given by 

\beq
     \frac{dQ}{dt} = \rho v A = \rho c \,(4 \pi r^2)
\eeq
where $r$ is the distance from the center.  This yields the formula for 
electric charge density in the galaxy

\beq
    \rho = \frac{\dot{Q}}{4 \pi c \,r^2} = \frac{\chi}{4\pi r^2}
\eeq
An electrostatic field satisfies the equation $ \nabla \cdot E = -\nabla^2 \phi = 4 \pi \rho $.
In the present case, 

\beq
     \frac{1}{r^2} \frac{d}{dr} \Big(r^2 \,\frac{d \phi}{dr}\Big) = -\frac{\chi}{r^2}
\eeq
Integrate once to obtain 

\beq
     E = -\frac{d \phi}{dr} = \frac{\chi}{r} - \frac{k}{r^2}
\eeq 
where $k$ is the constant of integration.  In a steady state, the central bulge (radius $r_b$) 
remains electrically neutral, so that $ k = \chi r_b $ and 

\beq
     E = \frac{\chi}{r} \Big(1 - \frac{r_b}{r}\Big)
\eeq

%\newpage
\section*{\large {\bf 3. The Gravitational Field.}}

Einstein's gravitational field equations are given by 

\beq
      \kappa R_{\mu\nu} + (T_{\mu\nu} - \frac{1}{2} g_{\mu\nu} T)  = 0
\eeq
where $\kappa = c^4/8\pi G.\,$  $T_{\mu\nu}$ is the energy tensor of matter
and electromagnetism.  Since the field is weak, it will be sufficient to 
calculate the single component 

\beq
     g_{00} = 1 + \frac{2}{c^2} \psi
\eeq
Moreover, the field is time-independent, in which case 
$c^2 R_{00} = - \nabla^2 \psi$. [2-4] Therefore, the gravitational potential 
satisfies a generalization of Newton's law 

\beq
    \nabla^2 \psi = \frac{8\pi G}{c^2} \Big(T_{00}- \frac{1}{2} \eta_{00} T \Big)
\eeq
The energy density of an electrostatic field is $T^{(e-m)}_{00} = E^2/8\pi$, while for matter
at rest $T^{(m)}_{00} = \rho_m c^2$.  The scalar $T^{(e-m)}$ is identically zero, while 
$T^{(m)} = \rho_m c^2$, and it follows that 

\beq
    \nabla^2 \psi = \frac{4\pi G}{c^2} \Big(\rho_m c^2 + \frac{E^2}{4\pi}\Big)
\eeq

%\newpage

\section*{\large {\bf 4. Gravity as $ r \longrightarrow \infty $.}}

In regions far from the galactic center, beyond the core and in the halo, 
the only source of gravitation is electric field energy.  In this region,
the gravitational potential satisfies the equation (5, 9)

\beq
     \frac{1}{r^2} \frac{d}{dr} \Big(r^2 \,\frac{d \psi}{dr}\Big) 
         = \frac{G}{c^2} \Big( \frac{\chi}{r} \Big)^2 \Big(1 - \frac{r_b}{r}\Big)^2
\eeq
Integrate this equation to find 

\beq
   \frac{d \psi}{dr} =  \frac{\chi^2 G}{c^2} \left\{\frac{1}{r}
                 - \frac{r_b}{r^2}\Big[2 \ln \frac{r}{r_b} + \frac{r_b}{r}\Big]\right\} + \frac{K}{r^2}
\eeq
The corresponding velocity (squared) curve is obtained by equating $ -d\psi/dr $ to the centripetal 
acceleration $ - v^2/r $

\beq
     v^2 = \frac{\chi^2 G}{c^2} \left\{1
                 - \frac{r_b}{r}\Big[2 \ln \frac{r}{r_b} + \frac{r_b}{r}\Big]\right\} + \frac{K}{r}
\eeq
In the limit $ r \longrightarrow \infty $, this yields the expression 

\beq
       v_\infty = \frac{\chi G^{1/2}}{c}
\eeq
As $ r \longrightarrow \infty $, the rotation curve gradually rises toward this 
asymptotic value.

%\newpage
\section*{\large {\bf 5. Gravity in the Core and Halo.}}

In the core of the galaxy, outside the dense central bulge, the mass density is assumed
to be a constant $\rho_c$.  In this region, the gravitational field satisfies

\beq
     \frac{1}{r^2} \frac{d}{dr} \Big(r^2 \,\frac{d \psi}{dr}\Big) 
         = \frac{4 \pi G}{c^2} \Big( \rho_c c^2 + \frac{E^2}{4 \pi} \Big)
\eeq
This equation integrates to 

\beq
   \frac{d \psi}{dr} =  \frac{4}{3} \pi \rho_c G r + \frac{\chi^2 G}{c^2} \left\{\frac{1}{r}
                 - \frac{r_b}{r^2}\Big[2 \ln \frac{r}{r_b} + \frac{r_b}{r}\Big]\right\} + \frac{K'}{r^2}
\eeq
Near the periphery of the bulge, the electrical term is zero, and 
$ d\psi/dr = GM_b/r^2_b $.  Therefore, the constant of integration is 

\beq
     K' = G M_b \Big(1 - \frac{\rho_c}{\rho_b}\Big)
\eeq
where $ M_b = \frac{4}{3} \pi \rho_b r^3_b $.  This yields the expression for velocities in the core

\beq
   v^2 = \frac{G M_b}{r}\Big(1 - \frac{\rho_c}{\rho_b}\Big) 
         + \frac{4}{3} \pi \rho_c G r^2 + \frac{\chi^2 G}{c^2} \left\{1
                 - \frac{r_b}{r}\Big[2 \ln \frac{r}{r_b} + \frac{r_b}{r}\Big]\right\} 
\eeq
The first term is a decreasing function of $r$, while the latter two terms are increasing.
At the edge of the core, where $ r = r_c $, formula (17) may be equated to (12).  This determines
the integration constant 

\beq
        K = G(M_b + M_c)
\eeq
where $ M_c = \frac{4}{3} \pi \rho_c (r^3_c -  r^3_b) $.  It follows that 
velocities in the halo satisfy

\beq
     v^2 = \frac{G(M_b + M_c)}{r} + \frac{\chi^2 G}{c^2} \left\{1
                 - \frac{r_b}{r}\Big[2 \ln \frac{r}{r_b} + \frac{r_b}{r}\Big]\right\} 
\eeq

%\newpage

\section*{\large {\bf 6. Numerical Results.} \rm [5, 6]}

The experimental value of $ v_\infty $ is very uncertain.  We choose
 $ v_\infty = 240 \;\mbox{\rm km} \; \mbox{\rm s}^{-1} $ in order  to obtain good
agreement in the core and halo.  Formula (13) then gives the value of $\chi$ for our galaxy

\beq
         \chi = 2.8 \times 10^{21} \; \mbox{statvolts}
\eeq
The mass $ M_b $ may be calculated from $ GM_b/r_b = v_b^2 $, using data from the rotation
curve: $ r_b = 600 \;\mbox{\rm parsecs (pc)}$ and $v_b = 260 \;\mbox{\rm km} \; \mbox{\rm s}^{-1} $ 

\beq
     M_b = 10^{10} \;\mbox{\rm solar masses}
\eeq
The mass $M_c$ is calculated by means of (19), using data for the Sun:
$ r_0 = 8.4 \;\mbox{\rm kpc}$ and $v_0 = 220 \;\mbox{\rm km} \; \mbox{\rm s}^{-1} $ 

\beq
     M_c = 1.5 \times 10^{10} \;\mbox{\rm solar masses}
\eeq
Together with the core radius, $ r_c = 4.0 \;\mbox{\rm kpc}$, this determines all parameters
in formulas (17) and (19).  They predict the following values for the rotation curve:
\vspace{.2in}

\begin{tabular*}{0.8\textwidth}{l|ccccccccccc}
  $r$    & .6&1&2&3&4&6&8.4&12&24&48&$\infty$\\
  $v$  &260&220&187&202&224&219&220&222&227&231&240
\end{tabular*}

\vspace{.2in}
\noindent The fit with the experimental curve is remarkable. [6]

The number density of protons may be calculated by setting $ \rho = n e $ in formula (2).
It varies from $ n = 10^{-13} \;\mbox{cm}^{-3} $ near the bulge 
to $ n = 10^{-16} \;\mbox{cm}^{-3} $ 
near the galactic fringe.  This is much less than
the density of cosmic rays that impinge upon Earth's atmosphere 
$ (10^{-10} \;\mbox{\rm particles}\;\mbox{\rm cm}^{-3})$.  Nevertheless, a large number of 
protons leave the halo each second:  set $ Q = Ne $ in (1) to find

\beq
    \frac{dN}{dt} = \frac{\chi c}{e} = 1.7 \times 10^{41} \;\mbox{\rm protons}\;\mbox{\rm sec}^{-1}
\eeq
(In a steady state, an equal number of electrons would leave the galaxy by way of the disk.)

The total energy of the electric field is given by 

\beq
    \int \frac{E^2}{8 \pi} \, 4\pi r^2 \, dr = 
                \frac{\chi^2 r_b}{2} \left\{\frac{r}{r_b} - 2 \ln \frac{r}{r_b} - \frac{r_b}{r}\right\}
\eeq
This integral diverges linearly.  If a cutoff is introduced at $ r = 60 \;\mbox{\rm kpc} $,
then the energy is equal to that of $3.5 \times 10^{11}$ solar masses.  This electric field energy is more than 10 times 
greater than the visible mass energy in our galaxy.  It is the `missing mass' of the galaxy.

The potential difference is given by

\beq
   \Delta \phi = -\int E \, dr  = -\chi \left\{\ln \frac{r}{r_b} + \frac{r_b}{r} - 1 \right\}
\eeq
Here, the divergence is logarithmic; with the cutoff, $ \Delta \phi = -3 \times 10^{24}\;\mbox{\rm volts}$.  
If each proton in (23) were to escape with $ 10^{24}\;\mbox{\rm eV} $, then the entire energy content 
of the galaxy would be depleted in $ 10^4 \;\mbox{\rm years} $. Therefore, the vast majority
of protons must attain an energy no greater than $ 10^{18} \;\mbox{\rm eV} $.  Cosmic ray measurements
show this to be the case.

\section*{\large {\bf 7. Concluding Remarks.}\footnote{The calculation of the electrostatic field
(section 2) was macroscopic, in that the charge density was assumed to be continuous.  In fact, 
the charge is due to relativistic protons moving through the galactic medium.  The density of 
interstellar plasma is estimated to be $.02 \;\mbox{\rm cm}^{-3}$, with a plasma frequency 
$\omega_p = 10^4 \;\mbox{\rm s}^{-1}$ and Debye screening length $\lambda_D = 5\times 10^3 
\;\mbox{\rm cm}$.  [7]    A proton passes through one Debye length in the time $c^{-1} \lambda_D =
1.7 \times 10^{-7} \;\mbox{\rm s}.$  This is much less than the response time of the plasma, 
$\omega_p^{-1} = 10^{-4} \;\mbox{\rm s}.$  Therefore, the plasma remains in equilibrium, and screening
does not occur.  It is transparent to the flow of proton charge.  Microscopically, the proton field
appears as a series of electromagnetic pulses. [8]  Macroscopically, it appears as an
electrostatic field.}}

The strength of the electric field (5) is not great.  It reaches a maximum of 
$ 0.4 \;\mbox{\rm statvolts}\;\mbox{\rm cm}^{-1} $ at $ r = 2r_b $, then decreases to 
$ 0.1 \;\mbox{\rm statvolts}\;\mbox{\rm cm}^{-1} $ near the galactic fringe.
(Galactic magnetic fields are of the order $10^{-6}$ gauss.  Thus,
the radial electric force on a given proton is far greater than any magnetic force.)
The energy density $ E^2/8\pi $ is similar in magnitude to the mass energy density
in the core $ \rho_c c^2 $.  The enormous total energy of the field
is due to its huge volume and not to its strength.  

This electric field is fixed in the galaxy, sustained by the flow
of highly energetic cosmic ray protons.  These protons, emerging from the center of 
the galaxy, are spatially ordered, because the speed of light yields a $ 1/r^2 $ 
steady state distribution of particles (2).  This large scale flow of cosmic ray protons differs
qualitatively from the thermal motion of particles that form the galactic plasma.  

The model yields two independent results which agree with experiment: (a) the
complete rotation curve of the galaxy and (b) its missing mass.  To the extent that spiral 
galaxies are alike, it also explains their contribution to the missing mass of the universe
as a whole.  Lastly, the model provides a new mechanism for the acceleration of galactic cosmic rays.

\newpage
\section*{\large {\bf References.}}

\begin{enumerate}
\item T. Ecrenaz, J. Bibring, M. Blanc, {\it The Solar System}, (Springer, 2003);
   cited in http://en.wikipedia.org/wiki/Solar\_wind
\item P.A.M. Dirac, {\it General Theory of Relativity}, (Wiley, 1975) section 16.
\item S. Weinberg, {\it Gravitation and Cosmology}, (Wiley, 1972) section 7.2.
\item C. Misner, K. Thorne, and J. Wheeler, {\it Gravitation}, (Freeman,
    \newline  New York, 1973) section 18.
\item B. Carroll and D. Ostlie, {\it An Introduction to Modern Astrophysics},
(Addison-Wesley, 1996) chapter 22. 
\item V. Kong and G. Rainey, {\it The Disk Rotation of the Milky Way Galaxy}, 
  http://www.csupomona.edu/\~{}jis/1999/kong.pdf 
\item R. Fitzpatrick, {\it Introduction to Plasma Physics}, \newline
 http://www.farside.ph.utexas.edu/teaching/plasma/plasma.html
\item J. Jackson, {\it Classical Electrodynamics}, (Wiley, 1998) 
  sect. 11.10.
\end{enumerate}

\end{document}